\documentclass[letterpaper,onecolumn]{article}
\usepackage{times}
\usepackage{graphicx}
\usepackage{amsmath}
\usepackage{amssymb}
\usepackage[top=2.0cm,left=2.0cm,right=2.0cm,bottom=2.0cm]{geometry}

\begin{document}

\date{}

\title{\Large\bf Blind Adaptive Successive Interference Cancellation for Multicarrier DS-CDMA}

\author{\small  Indu Shakya, Falah H.Ali, Elias Stipidis}
\maketitle \baselineskip=2 \baselineskip

\noindent\emph{\textbf{Abstract:}} A new adaptive receiver design
for the Multicarrier (MC) DS-CDMA is proposed employing successive
interference cancellation (SIC) architecture. One of the main
problems limiting the performance of SIC in MC DS-CDMA is the
imperfect estimation of multiple access interference (MAI), and
hence, the limited frequency diversity gain achieved in multipath
fading channels. In this paper, we design a blind adaptive SIC with
new multiple access interference suppression capability implemented
within despreading process to improve both detection and
cancellation processes. Furthermore, dynamic scaling factors derived
from the despreader weights are used for interference cancellation
process. This method applied on each subcarrier is followed by
maximum ratio or equal gain combining to fully exploit the frequency
diversity inherent in the multicarrier CDMA systems. It is shown
that this way of MAI estimation on individual subcarrier provides
significantly improved performance for a MC DS-CDMA system compared
to that with conventional matched filter (MF) and SIC techniques at
a little added complexity. Performance evaluation under severe
nearfar, fading correlation and system loading conditions are
carried out to affirm the gain of the proposed adaptive receiver
design approach.

\footnote{Submitted to Elsevier Computer Communications} \pagebreak
\noindent\section{Introduction}


CDMA is one of the promising multiple access schemes and is
currently being used for the third generation cellular wireless
systems. Future wireless communications systems need to meet the
increasing demands for highly flexible and efficient multiple access
techniques with improved performance, robustness to fading and other
interference \cite{hanzo_2003}, \cite{mc_ubiquotous}. Due to the
broadband nature of the transmission schemes used for such systems,
the propagation environment often introduces
inter-symbol-interference due to the large delay spread of multipath
fading channels. It is well known that Multicarrier CDMA is a very
effective scheme that mitigates the effects of multipath and also
provides frequency diversity that is inherent in wideband CDMA
channels \cite{mc_cdma_linnartz}-\cite{yang_zf_mmse_mc_dscdma}. The
main idea behind the Multicarrier CDMA is to use parallel
transmission of the same data over multiple frequency flat
subcarriers to avoid the multipath and hence to form a more robust
system in broadband environments.

A multicarrier scheme for CDMA that uses frequency domain spreading
also known as MC-CDMA was first introduced in
\cite{mc_cdma_linnartz}. Since then, a number of different variants
of Multicarrier CDMA schemes have been proposed; studies of which
can be found for example in \cite{mc_cdma_intro}-
\cite{mc_dscdma_rake}. In \cite{mc_cdma_intro}, the authors
described different multicarrier CDMA schemes both for the uplink
and downlink along with their relative performance advantages and
tradeoffs involved. A very useful survey of multicarrier CDMA
schemes for ubiquitous broadband wireless communications is carried
out by Yang and Hanzo in \cite{mc_ubiquotous}. Where it is found
that for a range of practical communications scenarios the
propagation channels may often become highly correlated in frequency
domain and thus MC-CDMA that uses spreading purely in the frequency
domain may not achieve the best performance. In this regard another
multicarrier CDMA scheme referred to as Multicarrier Direct Sequence
CDMA or MC DS-CDMA which is a combination of both MC-CDMA and
DS-CDMA can prove to be a very effective. The performance of MC
DS-CDMA is well investigated under different channel conditions and
detection schemes as can be found in \cite{mc_ubiquotous},
\cite{kondo_mcdscdma}, \cite{mc_dscdma_rake}, \cite{apic_mc_dscdma},
\cite{yang_zf_mmse_mc_dscdma}. The performance of MC DS-CDMA using
single user detection techniques are investigated and compared to
that with single carrier DS-CDMA in \cite{mc_dscdma_rake},
\cite{kondo_mcdscdma}. It is shown that the MC DS-CDMA offers more
improved performance under the same systems settings. Adaptive
filtering based receivers for multicarrier CDMA are also
investigated for example in \cite{cma_blind_mccdma},
\cite{adpt_mc_cdma_cma_qrd_rls}.

It is well known that single user or conventional MF based detection
scheme is suboptimal in multiuser settings as the interference
arising from other users are simply treated as background noise.
Multiuser detection \cite{verdu_1998} is a promising technique and
has been the subject of intense study in the last two decades for
single carrier CDMA. The optimum receiver using maximum likelihood
sequence estimation proposed by Verdu is known to offer substantial
gain in terms of both capacity and error performance. However
exponential computational complexity of the receiver has motivated
the research for suboptimal but lower complexity receivers. A very
useful simulation study of these suboptimal receivers have been
carried out in \cite{buehrer_simulationcomparison}. Among these
techniques, the Interference Cancellation (IC) subclass of receivers
have shown to offer both low complexity, high flexibility compared
with linear decorrelation based counterparts as studied by Andrews
in \cite{andrews_2005}. The IC receivers are further categorized
into successive (SIC) \cite{patel_1994}, \cite{bergman_lsic},
\cite{shakya_sic2007} and parallel (PIC) \cite{varanasi_1990},
\cite{yoon_ic}, \cite{divsalar_1998} techniques.

The IC techniques that were originally designed for single carrier
CDMA have also recently emerged as an effective approach to improve
the performance of multicarrier CDMA such as the work in
\cite{apic_mc_dscdma}, \cite{manohar_pic_mcdscdma},
\cite{fang_sic_mc_dscdma}, \cite{andrews_mc_sic}. In
\cite{fang_sic_mc_dscdma} and \cite{andrews_mc_sic} SIC technique is
investigated for the MC-CDMA and MC DS-CDMA, respectively. Also
recently, a weighted linear PIC for MC DS-CDMA is investigated in
\cite{manohar_pic_mcdscdma}. A well known problem with IC receivers
is the imperfect MAI estimation that arises due to the bias in
estimation when conventional correlators or MFs are used for this
purpose. This problem is addressed for single carrier CDMA by using
partial rather than full cancellation of the interference in
\cite{divsalar_1998}, \cite{bergman_lsic}. There are also adaptive
IC schemes that address this problem such as the LMS-PIC in
\cite{xue_1999} proposed for single carrier CDMA. However there has
been relatively less work on the adaptive IC schemes for
multicarrier CDMA systems. In \cite{apic_mc_dscdma} an adaptive
multistage PIC receiver for MC DS-CDMA using the LMS algorithm is
investigated. The LMS based adaptive SIC technique for the single
carrier CDMA also exists e.g. in \cite{asic_dscdma}. These schemes
use hard decision made on each stage to regenerate MAI estimates of
each user for the cancellation. To the best knowledge of the authors
adaptive approach to SIC for MC DS-CDMA has not yet been proposed.
In addition, we address the adaptive process using blind approach
and combined within the despreading process such that it also
provides additional multiple access interference suppression.
Recently, new techniques of coded transmission along with iterative
joint decoding and interference cancellation are becoming more and
more widespread such as \cite{pic_coding} and \cite{sic_siso}. Our
focus on this paper has been to propose a simple and low complexity
technique that improves the performance of the receiver from it's
design itself without using any coding and complex iterative
decoding, although they can be easily incorporated within the
proposed system.

Therefore, in this paper, we propose a new adaptive SIC receiver
design for MC DS-CDMA employing a simple CM algorithm within the
despreading process. Rather than using the hard decision of user
data, scaled soft output signals are used for a better estimation of
MAI for the cancellation, where the same despreader weights are used
for obtaining the dynamic scaling factors. Improved estimate of user
signals on each subcarrier are combined using maximum ratio (MRC) or
equal gain combining (EGC) method to maximize the frequency
diversity gain from the system. We also consider the effects of
problems such as nearfar and fading correlation that are commonly
encountered in practice. The system performance under different step
sizes of the adaptive algorithm is also investigated to reveal that
the proposed receiver can be further robustified under high system
loading condition with the choice of an optimum step size.

The paper is organized as follows. In section 2, the system model
and principle of conventional SIC technique are described. The main
ideas and detection algorithm of the proposed adaptive receiver are
presented in section 3. The performance results and comparisons with
other techniques are shown in section 4. Finally the paper is
concluded in section 5.\\\\

\noindent\section{System Description}
\noindent \subsection{System
Model} A $K$-user synchronous MC CD-CDMA system is considered. The
transmitted signal for the $k^{th}$ is given by
\begin{equation}
\label{eqn:tx_signal}
s_k(t)=\sqrt{\frac{2P}{L}}b_k(t)c_k(t)\sum_{l=1}^{L}\cos\{\omega_{l}t+\theta_{k,l}\}
\end{equation}
where $P$ is the signal power, $L$ is the number of subcarriers,
$b_k(t)=\sum_{m=-\infty}^\infty b_k(m)p_b(t-mT_b)$ is the data
signal, where $b_k(m)$ is a binary sequence taking values $[-1,+1]$
with equal probabilities, and $p_b(t)$ is the rectangular pulse
shaping function of period $T_b$. The spreading sequence is denoted
as $c_k(t)=\sum_{n=0}^{N-1} c_k(n)p_c(t-nT_c)$ with antipodal chips
$c_k(n)$ of pulse shaping function $p_c(t)$, period $T_c$, and
normalized power over a symbol period is equal to unity
$\int_{0}^{T_{b}}c_k(t)^2dt=1$. The spreading factor is $N=T_b/T_c$.
Finally, $\omega_{l}$ and $\theta_{k,l}$ are the $l^{th}$ subcarrier
frequency in radian/sec and initial carrier phase of the $k^{th}$
user uniformly distributed over $\{0,2\pi\}$, respectively.

In the proposed MC DS-CDMA system, we assume that the channel of
each subcarrier is a slowly varying frequency-nonselective Rayleigh
channel which remain constant over at least one symbol period. The
complex gain of the channel is given by
$\beta_{k,l}=g_{k,l}\exp^{j\phi_{k,l}}$, where $g_{k,l}$ is the
amplitude with zero mean and unit variance and $\phi_{k,l}$ is the
phase uniformly distributed over $\{0,2\pi\}$. We assume that the
channels of subcarriers have identical but not necessarily
independent distributions. The normalized channel correlation
$\rho_{l_{1},l_{2}}(\tau)$ can be given by \cite{jakes_mobile}
\begin{equation}
\label{eqn:tx_corr} \rho_{l_{1},l_{2}}(\tau)\triangleq
E[\beta^{*}_{k,l_1}(t)\beta_{k,l_2}(t+\tau)]=\frac{J_{0}(2\pi f_D
\tau)}{\sqrt{1+\Big(\frac{\Delta f}{\Delta f_c}\Big)^{2}}}
\end{equation}
where $E[.]$ is the expectation operator, $\beta^{*}$ is the complex
conjugation of $\beta$, $\Delta f= \mid f_{l_{1}}- f_{l_{2}}\mid$,
$\Delta f_c$ is the coherence bandwidth of the channel, $J_0(.)$ is
the zero order Bessel function of the first kind and $f_D$ is the
Doppler frequency.

The received signal is given by
\begin{equation}
\label{eqn:rx_signal}
r(t)=\sum_{k=1}^{K}\sqrt{\frac{2P}{L}}b_k(t)c_k(t)\sum_{l=1}^{L}\cos\{\omega_{l}t+\theta'_{k,l}\}+n(t)
\end{equation}
where $\theta'_{k,l}=\theta_{k,l}+\phi_{k,l}$ and $n(t)$ is the AWGN
with two sided power spectral density of $N_0/2$.

We assume that the system is fully synchronized and that the carrier
and fading channel phases are known perfectly at the receiver. It is
important to note that imperfect estimation of these parameters and
their effect on the system performance are important practical
issues as considered in \cite{fang_sic_mc_dscdma}. A downconverted
(baseband) equivalent of the received signal on $l^{th}$ subcarrier
$r^{l}(n)$ is obtained as follows
\begin{equation}
\label{eqn:rx_basesignal}
r^{l}(n)=LPF\Big\{\int_{nT_c}^{(n+1)T_c}\{r^{l}(t)\}'\cos\{\omega_{l}t+\theta'_{k,l}\}dt\Big\}
\end{equation}
where $\{r^{l}(t)\}'$ is the signal obtained after $r(t)$ is passed
through $l^{th}$ subcarrier bandpass filter. Furthermore, to
simplify the presentation in the description of different receiver
schemes, we use real valued signals for subsequent processes (
extension to the case of complex valued or higher modulation signals
can be straightforward ).\\

\noindent\subsection{Conventional SIC (CSIC) Receiver for MC
DS-CDMA}

In the conventional SIC as used in \cite{fang_sic_mc_dscdma} for
multicarrier DS-CDMA, the detection of user signals are performed
using MFs' or correlators' outputs. First the estimation of the
desired user is carried out, followed by the cancellation of its MAI
contribution from the remaining composite received signal. For this
purpose ordering is performed on the MF output signals to give the
strongest user say $k$ that gives the maximum of $L$ combined signal
output, given by
\begin{equation}
\label{eqn:mai_compo_sign1}
z^{l}_{k}=\max\Bigg\{{\sum_{l=1}^{L}\sum_{n=0}^{N-1}r^{l}(n)c_i(n)}\Bigg\},
1\leq i \leq K
\end{equation}
The estimate of $k^{th}$ user's data is taken as
\begin{equation}
\label{eqn:mai_compo_sign2} \hat{b}_k=sgn\Big
[\sum_{l=1}^{L}z^{l}_{k}\lambda^{l}_k\Big ]
\end{equation}
where, $sgn[.]$ denotes a sign function. As noted above in
(\ref{eqn:mai_compo_sign2}), the final decision is taken on the sum
of signals over $L$ subcarriers weighted by the combining weights
$\lambda^{l}_k$. There are two main combining methods used in the
literature \cite{hanzo_2003}, \cite{mc_cdma_intro},
\cite{mc_cdma_linnartz}: maximum ratio combining and equal gain
combining. The weights are taken as $\lambda^{l}_k=g^{l}_k, \forall
k, \forall l$ for the MRC and $\lambda^{l}_k=1, \forall k, \forall
l$ in the case of EGC.

After the decision of the $k^{th}$ user, it's signal $z^{l}_k$ is
respread using its spreading sequence
$\textbf{c}_k=[c_k(1),c_k(2),..c_k(N)]^{T}$ where $\{.\}^{T}$
denotes a transpose operation and subtracted from
$\textbf{r}^{l}_k=[r^{l}_k(1),r^{l}_k(2),..,r^{l}_k(N)]^{T}$ to form
received signal for the next strongest user $\textbf{r}^{l}_{k+1}$
as follows
\begin{equation}
\label{eqn:mai_compo_sign3}
\textbf{r}^{l}_{k+1}=\textbf{r}^{l}_{k}-z^{l}_k(m)\textbf{c}_k
\end{equation}
The processes (\ref{eqn:mai_compo_sign1}),
(\ref{eqn:mai_compo_sign2}) and (\ref{eqn:mai_compo_sign3}) are
carried out on all $L$ subcarrier branches until all user's data are
detected.

\noindent\section{Proposed Blind Adaptive SIC (ASIC) Receiver}

\noindent\subsection{Main Ideas and Design Architecture} The
proposed receiver addresses the discussed problems with it's unique
design as follows. \textbf{Firstly}, instead of only correlating the
input signal with a user specific sequence, adaptive despreading is
used utilizing the constant modulus property of the desired user's
transmitted signal to reduce the variance of MAI, that is known to
be zero mean and approximately Gaussian distributed signal. It also
has a desirable property of generating the adaptive weights that
does not allow a decision statistic signal to revert it's sign when
the presence of MAI tends to do so. The adaptive algorithm is based
on CMA, which is a simple blind algorithm commonly used in the
literature such as \cite{cma_review}, \cite{lee_1996} that tries to
maintain constant modulus of the signals at the output. It is used
here in a different way within the despreader to reduce MAI where
the weights are also used to obtain refined MAI estimates for the
cancellation. The complexity of the algorithm is only $O(N)$
computation per symbol per user, where $N$ is the length of the
weight vector. Provided that the algorithm is fast enough to track
the changes in MAI power variations with optimum weights, the
decision error due the MAI effects can be completely avoided.
However, this is very difficult to achieve in practice since the CM
algorithm may always have some inevitable misconvergence problems.
\textbf{Secondly}, to address this problem, the adaptive despreader
weights are used to obtain dynamic scaling factors that are suitably
implemented within the SIC stages. It is intuitive that the more MAI
is suppressed during the despreading, the more reliable is it's
cancellation. \textbf{Finally}, the improved estimates of user's
signals obtained on the individual subcarrier are combined using EGC
or MRC method to fully exploit the frequency diversity inherent in
MC CDMA systems. Our results show that the adaptive approach to the
estimation and cancellation minimizes the effect of MAI even under
high system loading and severe nearfar conditions.

\begin{figure}[htbp]
\begin{center}
\includegraphics[width=6.6 in]{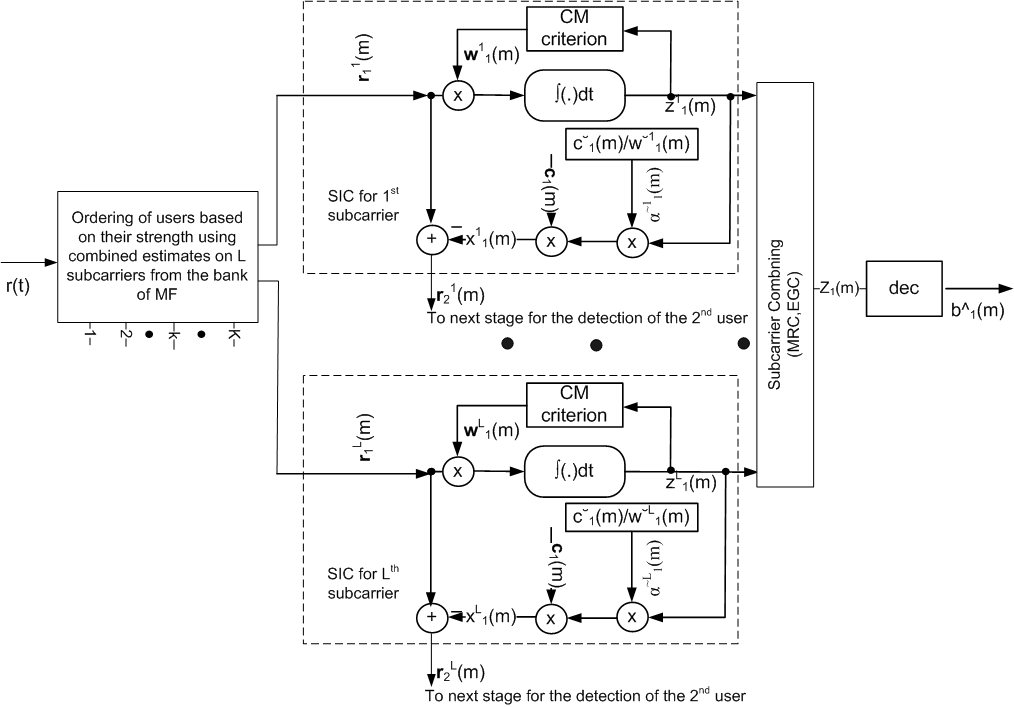}
\caption{Proposed Adaptive SIC receiver design for the detection of
the first user} \label{fig:asic}
\end{center}
\end{figure}

\begin{figure}[htbp]
\centering
\includegraphics[width=4.80 in]{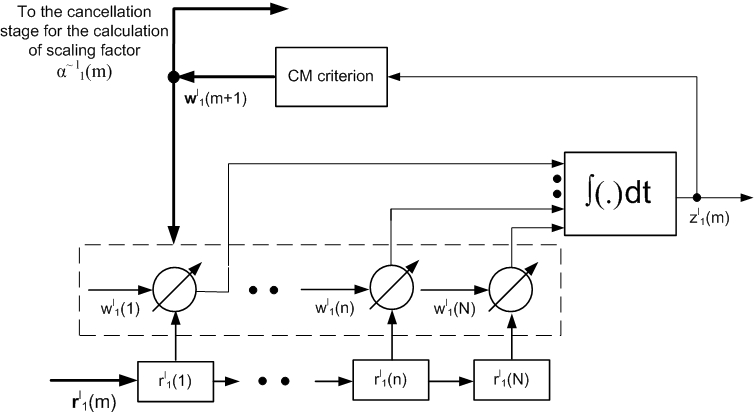}
\caption{Adaptive despreading and generation of weights for
obtaining the scaling factor for the first user}
\label{fig_cma_despread}
\end{figure}

The proposed SIC architecture block diagram is shown in Figure 1 for
the first user. The same processes are applied to all $K$ stages. In
this architecture, the effect of strong users (interferers) are
removed at each successive stage, which aids the detection and
cancellation for the weaker user. At every symbol period, the
received signal $r(t)$ is down converted to $L$ baseband equivalent
of subcarriers signals and sampled at the chip rate to form the
vector $\textbf {r}^{l}(m), 1\leq l \leq L$ each of length $N$
chips. Bank of MF output based power sorter is employed to order
user signals according to their combined strength on $L$
subcarriers. The strongest user signal is then selected for the
first stage for the estimation and detection of it's transmitted
data. Adaptive algorithm based on CM criterion embedded within the
despreader is used to adjust the amplitude of incoming signal at the
chip rate. For example, the despreading process for the received
signal at the first (strongest) user stage on $l^{th}$ subcarrier is
shown in Figure 2. The output $z^{l}_1(m)$ is also weighted by a
scaling factor $\tilde{\alpha}^{l}_1(m)$ utilizing the despreader
weights $\textbf{w}^{l}_1(m)$, spread with $\textbf{c}_1(m)$ and
subtracted from the received signal $\textbf{r}^{l}_1(m)$ to form
the input to the next stage for the second strongest user. This
process is repeated for each user until the weakest user is
detected.

\noindent\subsection{Detection Algorithm} At the first symbol
period, the weights of the adaptive despreaders are initialized with
user's spreading sequence $\textbf w^{l}_k(1)=\textbf c_k(1)$.
Without loss of generality, it is assumed that the first user
(strongest among K users) to be detected is user $1$. Similarly next
strongest user is assigned an index as user 2 and so on. At the
first stage, the received signal on the $l^{th}$ subcarrier can be
expressed as $\textbf r^{l}(m)=\textbf r^{l}_1(m)$. The remaining
composite signal after $k$ cancellation stages is expressed as
$\textbf r^{l}_{k+1}(m)$.

At $k^{th}$ detection stage, the decision statistic $z^{l}_k(m)$ is
obtained by multiplying chips of $\textbf {r}^{l}_k(m)$ with the
vector of weights $\textbf w^{l}_k(m)$ and summed over the symbol
period given by
\begin{equation}
\label{eqn:decision_stat1}
z^{l}_k(m)=\Big\{\textbf {w}^{l}_k(m)\Big\}^{T} \textbf r^{l}_{k}(m)
\end{equation}
The CM criterion $J_{CM}$ can be written as minimization of the
following cost function
\begin{equation}
\label{eqn:decision_stat2}
J_{CM}=E\Big\{{z^{l}_k(m)^2}-\gamma\Big\}^2
\end{equation}
where $\gamma$ is the dispersion constant, which is equal to unity
for BPSK signals. The instantaneous error signal $e^{l}_k(m)$ is
calculated as
\begin{equation}
\label{eqn:bpsk_sign} e^{l}_k(m)=z^{l}_k(m)\Big\{z^{l}_k(m)^2-\gamma
\Big\}
\end{equation}
The estimated gradient vector of the error signal is then calculated
by
\begin{equation}
\label{eqn:grad_vector_error} \nabla^{l}_k(m)= \textbf
r^{l}_{k}(m)e^{l}_k(m)
\end{equation}
Using the gradient of (\ref{eqn:grad_vector_error}), the weight
vector at next symbol $\textbf w^{l}_k(m+1)$ is updated as follows
\begin{equation}
\label{eqn:weight_vector} \textbf{w}^{l}_k(m+1)=\textbf
w^{l}_k(m)-\mu\nabla^{l}_k(m)
\end{equation}
where, $\mu$ is the step-size used for adapting the elements of a
weight vector to minimize the cost function
(\ref{eqn:decision_stat2}).

Next, the cancellation process requires amplitude estimate
$g^{l}_k(m)$ of the detected user's signal along with it's spreading
sequence $\textbf{c}_k(m)$. To achieve an estimate as close to
$g^{l}_k(m)$ as possible, first a scaling factor
$\tilde{\alpha}^{l}_k(m)$ is obtained using the despreader weights
and the known spreading sequence of the user $\textbf{c}_k(m)$ as
follows
\begin{equation}
\label{eqn:est_scaling_factor1} \tilde{\alpha}^{l}_k(m)=\frac
{\breve{c}_k(m)}{\breve{w}^{l}_k(m)}
\end{equation}
where, $\breve{c}_k(m)$ and $\breve{w}^{l}_k(m)$ are the mean
amplitude of chips of user's spreading sequence and elements of the
weight vector updated by the CM algorithm, respectively and are
given in (\ref{eqn:est_scaling_factor2}) and
(\ref{eqn:est_scaling_factor3}) below
\begin{equation}
\label{eqn:est_scaling_factor2}
\breve{c}_k(m)=\frac{1}{N}\sum_{n=0}^{N-1} \Big|
c_k\big\{(m-1)N+n\big\}\Big|
\end{equation}

\begin{equation}
\label{eqn:est_scaling_factor3}
\breve{w}^{l}_k(m)=\frac{1}{N}\sum_{n=0}^{N-1}
\Big|w^{l}_k\big\{(m-1)N+n\big\}\Big|
\end{equation}
The soft estimate signal $z^{l}_k(m)$ is then scaled with its new
amplitude estimate $\tilde {\alpha}^{l}_k(m)$ and spread to generate
the cancellation term as follows
\begin{equation}
\label{eqn:gen_cancel_term}
\textbf{x}^{l}_k(m)=\tilde{\alpha}^{l}_k(m)
{z}^{l}_k(m)\textbf{c}_k(m)
\end{equation}
The remaining composite signal after the interference cancellation
is given by
\begin{equation}
\label{eqn:interference_cancel}
\textbf{r}^{l}_{k+1}(m)=\textbf{r}^{l}_{k}(m)-\textbf{x}^{l}_k(m);
\end{equation}
After collecting $L$ samples of $z^{l}_k(m)$, the signals are
delivered to the subcarrier combining stage to improve the final
estimate of the desired user's data. Each signal $z^{l}_k(m)$ is
weighted by a weight $\lambda^{l}_k(m)$ and then combined to form
the final decision variable $Z_k(m)$ as follows:
\begin{equation}
\label{eqn:combining} Z_k(m)=
\sum_{l=1}^{L}\lambda^{l}_k(m)z^{l}_k(m)
\end{equation}
As noted earlier, the combining weights here are taken as
$\lambda^{l}_k(m)=g^{l}_k(m), \forall k, \forall l$ for the MRC and
$\lambda^{l}_k(m)=1, \forall k, \forall l$ in the case of EGC.
Finally the decision making process performs hard decision to obtain
the data as follows
\begin{equation}
\label{eqn:decision_process} \hat{b}_k(m)=dec\{Z_k(m)\}
\end{equation}
where $dec\{.\}$ is a simple sign detector for BPSK signals.

The processes shown in
(\ref{eqn:decision_stat1})-(\ref{eqn:decision_process}) are repeated
for each detection stage until the weakest user is detected.

We summarize the detection algorithm in Table 1 below for a more
concise presentation.\\

\begin{center}
\begin{tabular}{l l}\\
\hline
{Initialize the adaptive algorithm and select the step size $\mu$} \\
{At $m=1$, set $\textbf w^{l}_k(1)= \textbf c_k(1)$, $\forall k$, $\forall l$} \\
\hline
{for $m=1,2,..$}\\
{\quad Sort users based on their strength in descending order and store their indices $k$ }\\
{\quad \quad  for $k=1,2,..,K$} \\
{\quad \quad \quad for $l=1,2,..,L$} \\
{\quad \quad \quad \quad} 1. Despread $\textbf r^{l}_k(m)$ using the weight vector $\textbf w^{l}_k(m)$ and store the sample of $z^{l}_k(m)$ \\
{\quad \quad \quad \quad} 2. Evaluate the CM cost function $J_{CM}$ and calculate gradient vector, $\nabla^{l}_k(m) $ \\
{\quad \quad \quad \quad} 3. Update weights for next symbol, $\textbf w^{l}_k(m+1)$\\
{\quad \quad \quad \quad} 4. Calculate the scaling factor $\tilde\alpha^{l}_k(m)$ and regenerate the cancellation term $\textbf {x}^{l}_k(m)$ \\
{\quad \quad \quad \quad} 5. Cancel the signal $\textbf {x}^{l}_k(m)$ from the remaining total composite signal $\textbf{r}^{l}_{k}(m)$\\
{\quad \quad \quad end for} \\
{\quad \quad \quad} Perform MRC or EGC over $L$ samples of $z^{l}_k(m)$ to form $Z_k(m)$ and then obtain $\hat b_k(m)$ \\
{\quad \quad end for }\\
{end for}\\
\hline
\end{tabular}
\\
Table 1. The proposed ASIC receiver algorithm steps
\end{center}

\noindent\section{Performance Results and Comparisons}

\noindent \subsection {Assumptions}

A synchronous uplink MC DS-CDMA system of $K$ BPSK modulated users
with $L=2$ and coherent demodulation is assumed in all simulations.
Raised cosine pulse shaping filter with roll factor $\alpha=0.5$ is
used within each subcarrier. Equal power users $P_k=P, \forall k$
are assumed unless stated otherwise. Short binary Gold sequences of
length N = 31  are used for spreading users' data. The channel is
Rayleigh flat fading on each subcarrier with normalized Doppler rate
$f_{D}T_b=0.003$ and perfect channel estimation is assumed for the
case of MRC combing method. The generation of correlated fading of
subcarriers are obtained using the method proposed in
\cite{correlated_rayleigh} that is based on Jakes model
\cite{jakes_mobile}. A step size of $\mu_k=0.0001$, for all users is
assumed in the adaptive algorithm. The selection of step size is
generally based on the spreading factor, system loading and the
dynamic range of the received signal and also has direct effect on
the system performance as will be shown later.

\noindent \subsection {Numerical Results}

\noindent \textbf{BER vs. SNR:} The BER performance of the proposed
multicarrier blind Adaptive SIC (ASIC) is shown in Figure 3 for
system loading of $K=20$ users and $\rho_{l_1,l_2}=0$ using EGC and
MRC methods, denoted as ASIC-EGC and ASIC-MRC, respectively. For
comparison purposes the BER obtained using conventional SIC (CSIC)
and MF receivers using the two combining methods are also given. As
can be clearly seen from the figure, the proposed technique offers
significant improvement in the BER compared with the other
techniques for both cases of EGC and MRC. It approaches much closer
to the single user performance. It is also noted that the
performance with EGC and MRC are very similar for all receivers
under the system settings considered.

\begin{figure}[htbp]
\begin{center}
\includegraphics[width=5.0 in]{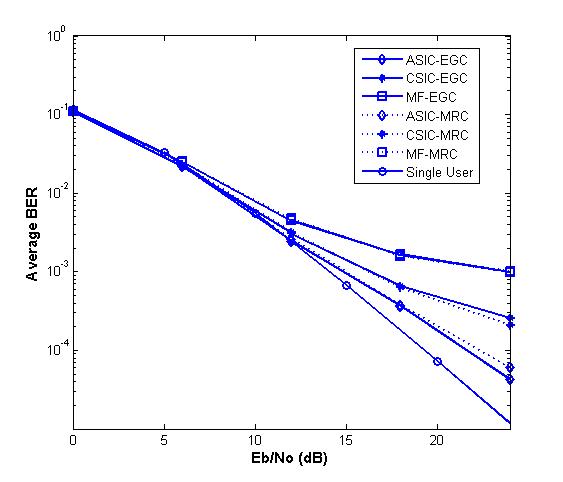}
\caption{BER of the Adaptive SIC receiver for MC DS-CDMA with $L=2$
in Rayleigh fading channels for $K=20$, $\rho_{l_1,l_2}=0$,
$\mu=0.0001$ and Gold sequences of $N=31$} \label{fig:asic_ber_snr}
\end{center}
\end{figure}

\noindent \textbf{User capacity:} In Figure 4, the user capacity of
ASIC is compared with CSIC and MF receivers in Rayleigh fading
channels with $\rho_{l_1,l_2}=0$ and $E_b/N_0=20$ dB. It is evident
that proposed ASIC-EGC and ASIC-MRC achieve much improved user
capacity compared with both CSIC and MF. For example, at the BER of
$2\times 10^{-4}$, the ASIC-EGC can support 20 users compared with
16 users and 8 users with CSIC-EGC and MF-EGC, respectively. Also it
is noted that the performance of the ASIC for both EGC and MRC are
very similar. The CSIC-MRC is shown to outperform CSIC-EGC under low
user loading conditions and the opposite for high loading
conditions. MF-EGC seems to perform better than MF-MRC for the whole
range of user loading conditions.

\begin{figure}[htbp]
\begin{center}
\includegraphics[width=5.0 in]{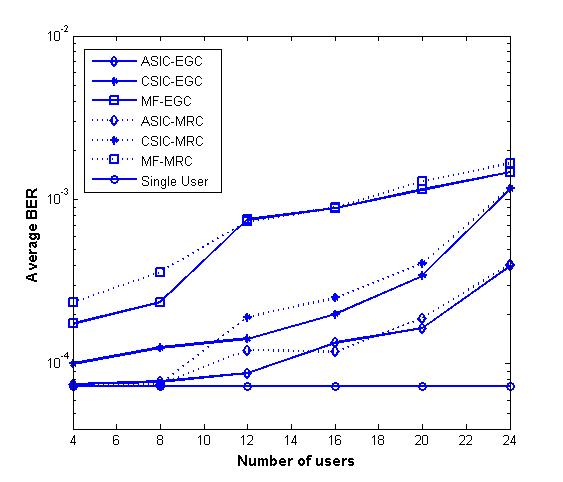}
\caption{User capacity performance of the Adaptive SIC receiver for
MC DS-CDMA with $L=2$ in Rayleigh fading channels under $E_b/N_0=20$
dB, $\rho_{l_1,l_2}=0$, $\mu=0.0001$ and Gold sequences of $N=31$}
\label{fig:asic_ber_capacity}
\end{center}
\end{figure}

\noindent \textbf{Effects of nearfar conditions:} To further affirm
the performance gains of the proposed ASIC receiver for MC DS-CDMA,
we also consider the system with different user nearfar conditions.
Firstly, in Figure 5, we assess and compare the user capacity
performance of receivers under the nearfar ratio of $\Omega=10$ dB.
This is obtained by letting the desired user which is also the
weakest one to have unity power $P_k=1$ whereas all the other users
$i,i\neq k$ have their power uniformly distributed within the range
$P_i\in\{0,10\}$. The step size used in the adaptive algorithm of
the proposed receiver is set as $\mu=\frac{0.0001}{\Omega}$ to
compensate for the nearfar situation. The other system settings are
the same as the previous equal power case. As can be noted from the
figure, the ASIC-MRC shows much improved performance compared with
CSIC-MRC and MF-MRC under various user loading conditions. As
expected the MF-MRC shows rapid degradation in the BER as the user
loading increases. The CSIC-MRC also shows improved performance
compared with MF-MRC but performs much worse than the ASIC-MRC under
high loading conditions.

\begin{figure}[htbp]
\begin{center}
\includegraphics[width=5.0 in]{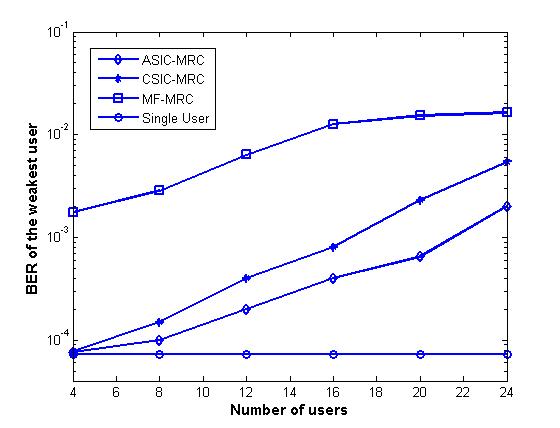}
\caption{Performance of the Adaptive SIC receiver with MRC with
$L=2$ for MC DS-CDMA in nearfar condition with $\Omega=10$ dB  and
with desired (weakest) user's $E_b/N_0=20$ dB in Rayleigh fading
channels with $\rho_{l_1,l_2}=0$, $\mu=0.00001$ and Gold sequences
of $N=31$} \label{fig:asic_ber_k4_24nearfar10}
\end{center}
\end{figure}

Furthermore, in Figure 6 the receivers are investigated with a
system consisting of $K=20$ users and under different nearfar
conditions of $\Omega\in \{0,20\}$ dB. The step size of the adaptive
algorithm used in the proposed receiver is varied according to the
nearfar condition given by $\mu=\frac{0.0001}{\Omega}$. The ASIC-MRC
continues to show improved performance even under high values of
$\Omega$. The performance gap between the ASIC and CSIC becomes
wider in this case showing impressive performance advantage of the
proposed receiver design approach. This result can also be
attributed to the improved detection and MAI cancellation processes
of the ASIC giving it more resilience under the nearfar conditions.

\noindent \textbf{Effects of step-size:} The step size is an
important parameter of the adaptive algorithm
\cite{haykin_adpt_filter} with direct effects on the convergence. In
Figure 7, we assess the impact of different step sizes on the BER
performance of the proposed receiver with MRC under $E_bN_0=20$ dB
for $K=10,16,20$ and $24$ users in Rayleigh fading channels with
$\rho_{l_1,l_2}=0$. The BER obtained using the step size of
$\mu=0.0001$ used in earlier simulations is also shown. It can be
seen that the choice of step size can have significant effect on the
system performance. Under low loading condition of $K=10$, the
performance is nearly identical for a range of step sizes. However
as K increases, the effect of step size has more significant impact
on the BER performance. Interestingly, it is noted for example for
$K=24$, the BER performance of the proposed adaptive receiver can be
further improved. With the choice of optimum step
$\mu_{opt}=0.0013$, BER performance as low as $8.5\times 10^{-5}$
can be achieved which is very near to the single user performance of
$6.75\times 10^{-5}$.

\begin{figure}[htbp]
\begin{center}
\includegraphics[width=5.0 in]{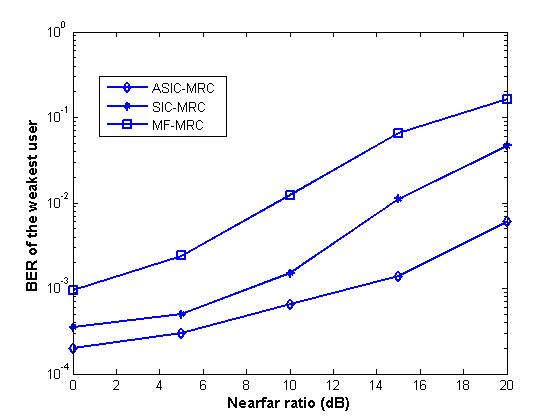}
\caption{Performance of the Adaptive SIC receiver with MRC for MC
DS-CDMA for $K=20$ users, $L=2$  in nearfar condition with
$\Omega\in\{0,20\}$ dB and with desired (weakest) user's
$E_b/N_0=20$ dB in Rayleigh fading channels with $\rho_{l_1,l_2}=0$,
$\mu=0.0001/\Omega$ and Gold sequences of $N=31$}
\label{fig:asic_ber_k20nearfar_0_20db}
\end{center}
\end{figure}

\begin{figure}[htbp]
\begin{center}
\includegraphics[width=5.0 in]{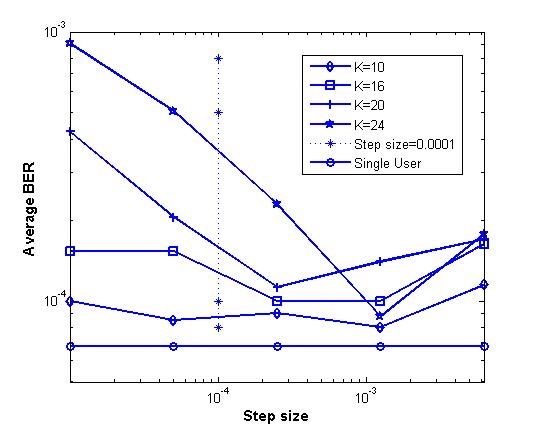}
\caption{Effects of step size on the performance of the Adaptive SIC
receiver with MRC for MC DS-CDMA with $L=2$ under $E_bN_0=20$ dB in
Rayleigh fading channels $\rho_{l_1,l_2}=0$ for $K=10, 16, 20, 24$
users and Gold sequences of $N=31$} \label{fig:asic_ber_step}
\end{center}
\end{figure}

\noindent \textbf{Effects of subcarrier correlation:} The results
obtained so far assumed independent fading across different
subcarrier channels. It is interesting to see the effect of the
fading correlation on the performance of different receivers for the
sake of more practical system design purposes. In Figure 8, we plot
the BER of ASIC-MRC under $E_b/N_0=20$ dB for $K=10$ and $20$ equal
power users and investigate the effect of different correlation
coefficients $\rho_{l_1,l_2}\in\{0,0.8\}$. It can be noted that the
BER performance of all receivers degrade gracefully as the magnitude
of $\rho_{l_1,l_2}$ increases. This result provides useful insight
on the performance of ASIC for operation in practical wireless
channel conditions where some correlation between the subcarriers
may always exist.

\begin{figure}[htbp]
\begin{center}
\includegraphics[width=5.0 in]{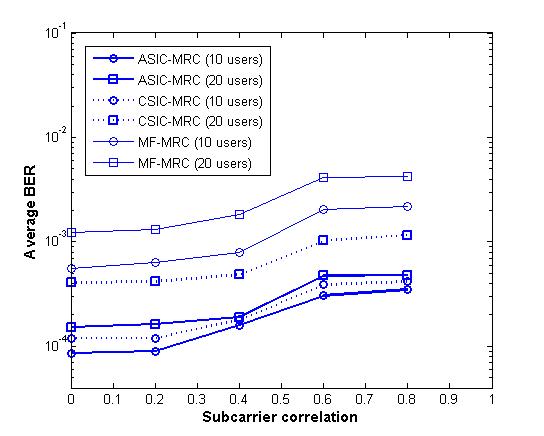}
\caption{Effects of subcarrier correlation $\rho_{l_1,l_2}$ on the
performance of the Adaptive SIC receiver with MRC for MC DS-CDMA
with $L=2$ in Rayleigh fading channels for $K=10$ and $20$ users and
Gold sequences of $N=31$} \label{fig:asic_ber_corr}
\end{center}
\end{figure}

\noindent \section {Conclusion}

We proposed and evaluated a new adaptive receiver using successive
interference cancellation method for MC DS-CDMA systems. The
proposed technique employs a simple cost to perform adaptive
despreading to weight each incoming chip signal in a blind manner to
provide the despreading function along with the additional multiple
access interference suppression capability. The same despreader
weights are also used to form a more reliable MAI estimate for the
cancellation stage. It is shown that this approach to multiuser
detection offers significant performance gain compared with
conventional SIC and MF detection techniques while retaining the low
receiver complexity. For example, it is shown to support 20 users
compared with 16 and 8 users with the conventional techniques. With
the use of optimum step size, even for a high loading of 24 users it
offered a BER of as low as $8.5\times 10^{-5}$, which is very near
to the single user performance. Furthermore, for a range of
simulation scenarios the results showed that the adaptive receiver
is also much more resilient to the near far problem. Our future work
will be to combine this technique with additional error correction
coding to further improve the system performance.

\end{document}